\documentclass[journal]{IEEEtran}
\IEEEoverridecommandlockouts

\usepackage{cite}
\usepackage{amsmath,amssymb,amsfonts}
\usepackage[ruled,linesnumbered]{algorithm2e}
\usepackage{graphicx}
\usepackage{textcomp}
\usepackage{xcolor}
\newtheorem{lemma}{\indent Lemma}
\def\BibTeX{{\rm B\kern-.05em{\sc i\kern-.025em b}\kern-.08em
    T\kern-.1667em\lower.7ex\hbox{E}\kern-.125emX}}
\usepackage[letterpaper, right=0.64in, left=0.63in, top=0.7in, bottom=1.04in]{geometry}
\begin{document}

\title{Beam-aware Kernelized Contextual Bandits for User Association and Beamforming in mmWave Vehicular Networks}

\author{\IEEEauthorblockN{
		Xiaoyang He\IEEEauthorrefmark{2} 
		and
		Manabu Tsukada\IEEEauthorrefmark{1}
	}

	\IEEEauthorblockA{\IEEEauthorrefmark{2}Graduate School of Electronics and Communication Engineering, Sun Yat-sen University, China}
	
	\IEEEauthorblockA{\IEEEauthorrefmark{1}Graduate School of Information Science and Technology, The University of Tokyo, Japan}

	\thanks{
	The work of He was supported in part by the scholarship from China Scholarship Council (CSC) under the Grant CSC 202406380226.
	The work of Tsukada was supported by JST ASPIRE Grant Number JPMJAP2325, Japan.
	}
	Email: hexy35@mail2.sysu.edu.cn, mtsukada@g.ecc.u-tokyo.ac.jp
}

\maketitle
\pagestyle{empty}
\thispagestyle{empty}

\begin{abstract}
Timely channel information is necessary for vehicles to determine both the serving base station (BS) and the beamforming vector, but frequent estimation of fast-fading mmWave channels incurs significant overhead.
To address this challenge, we propose a Beam-aware Kernelized Contextual Upper Confidence Bound (BKC-UCB) algorithm that estimates instantaneous transmission rates without additional channel measurements by exploiting historical contexts such as vehicle location and velocity, together with past observed transmission rates.
Specifically, BKC-UCB leverages kernel methods to capture the nonlinear relationship between context and transmission rate by mapping contexts into a reproducing kernel Hilbert space (RKHS), where linear learning becomes feasible.
Rather than treating each beam as an independent arm, the beam index is embedded into the context, enabling BKC-UCB to exploit correlations among beams to accelerate convergence.
Furthermore, an event-triggered information sharing mechanism is incorporated into BKC-UCB, enabling information exchange only when significant explorations are conducted to improve learning efficiency with limited communication overhead.
\end{abstract}


\section{Introduction}

To support high data rates, low latency, and massive connectivity, mmWave communication has emerged as a promising technology for vehicular networks \cite{9779354}.
Establishing reliable communication between vehicles and base stations (BSs), which involves user association and beamforming, is particularly challenging.
Conventional offline schemes rely on accurate channel state information (CSI) to optimize user association and compute beamforming vectors via methods such as singular value decomposition (SVD) \cite{6497017,8677293}.
Rapidly varying vehicular channels make timely CSI acquisition costly, rendering such offline schemes impractical.
The Multiple Signal Classification (MUSIC) algorithm enables beamforming without explicit CSI by estimating the direction of arrival (DOA) \cite{1143830}, but it requires precise array modeling and a large number of samples, which is infeasible for fast-fading mmWave vehicular channels with short coherence time \cite{al2022review}.

The multi-armed bandit (MAB) framework offers an effective online learning method to cope with dynamic channel conditions with low computation and communication overhead without knowledge of CSI.
In this framework, each vehicle can be regarded as a learning agent, while candidate BSs and available beamforming vectors are modeled as arms.
At each period, a vehicle selects a BS and a beamforming vector for communication, and receives the transmission rate as the reward.
However, in mmWave vehicular networks, the short wavelength and high mobility lead to highly dynamic channels, leading to non-stationary and non-i.i.d rewards.
To address this issue, the contextual MAB (CMAB) framework has been introduced.
Prior studies have applied the CMAB framework to user association \cite{8472783, CC-UCB} and beamforming \cite{li2020smart}.
These approaches partition the context space into multiple small regions, where each observed context-reward sample updates only a small number of regions.
However, in scenarios with a wide coverage area, the context space becomes excessively large, requiring substantial exploration and resulting in slow convergence.
In addition, existing studies typically treat beams as independent arms \cite{li2020smart}, which leads to slow convergence when a large codebook is adopted.

In this paper, we propose a Beam-aware Kernelized Contextual Upper Confidence Bound (BKC-UCB) algorithm for joint user association and beamforming in mmWave vehicular networks without relying on CSI.
The context is defined as vehicle location, velocity, the number of concurrent transmissions, and beam index.
Within a given context, similar beams, locations, velocities, and concurrent transmissions result in comparable channel characteristics and interference, so the transmission rate can be modeled as drawn from an i.i.d. reward distribution conditioned on that context.
Each context is implicitly projected into a Reproducing Kernel Hilbert Space (RKHS) through a feature map, where the expected reward is assumed to have a linear relationship with the mapped context.
Furthermore, instead of treating each available beam as an independent arm, the beam index is incorporated into the context as an additional dimension.
Under kernel methods, previously sampled beams can collaboratively contribute to the estimation for the current context, leading to faster convergence.
Information sharing is triggered only when significant explorations are conducted, accelerating convergence while maintaining low communication overhead.

The rest of this paper is organized as follows.  
Section II describes the system model, followed by the proposed BKC-UCB algorithm in Section III.  
Simulation results are presented in Section IV, and Section V concludes the paper.

\section{System Model}
We first introduce the vehicle mobility model, beamforming, channel models, and then formulate the joint optimization problem for user association and beamforming.

Consider a finite time horizon $T \in \mathbb{N}$, during which vehicles update their positions at each time period $t = 1, \dots, T$.  
At period $t$, the set of active vehicles and BSs is denoted by $\mathbb{U}(t)$ and $\mathbb{B}$, respectively.  
The number of newly arriving vehicles in each period is modeled by a Poisson process with intensity $\lambda$.  
Each BS is assumed to be equipped with a half-wavelength spaced Uniform Linear Array (ULA) consisting of $N_R$ antenna elements, whereas each vehicle deploys a ULA with $N_T$ antenna elements for communication.

We assume that both the transmitter and the receiver employ a hierarchical binary-tree structured codebooks \cite{xiao2016hierarchical}, denoted by $\mathcal{W}_T$ and $\mathcal{W}_R$, respectively.
Without loss of generality, we discuss the codebook for the transmitter side, i.e., the vehicle side.
Each beam in the codebook is specified by two parameters: the steering angle $\psi$ and a layer index $l \in 1,2,\dots, L_m$, where the maximum layer of the codebook is $L_m = \lceil\log_2 N_T\rceil$.
For a given steering angle, the layer determines the width of the main-lobe, and a large layer indicates a narrow beam.
The beamforming vector $\mathbf{w}(\psi, l)$ is defined as,
\begin{equation}
	\begin{aligned}
		\mathbf{w}(\psi, l) &= \left[ \mathbf{a}\left(\psi, N(l)\right)^{\mathrm{T}}, \mathbf{0}_{N_T - N(l)}^{\mathrm{T}} \right]^{\mathrm{T}}, 
		N(l) = \min(2^l, N_T), \\
		\mathbf{a}(\psi, N) &= \frac{1}{\sqrt{N}}\left[e^{j\pi 0\sin(\psi)}, \dots, e^{j\pi (N-1)\sin(\psi)}\right]^{\mathrm{T}}.
	\end{aligned}
\end{equation}

Let $\mathbf{w}^r_i(t) \in \mathcal{W}_R \subseteq \mathbb{C}^{N_R \times 1}$ and $\mathbf{w}^t_i(t) \in \mathcal{W}_T \subseteq \mathbb{C}^{N_T \times 1}$ denote the combining vector and beamforming vector between vehicle $i$ and BS $a$, respectively.
When BS $a$ serves vehicle $i$, the effective channel gain is expressed as
\begin{equation}
	\label{h_ia_0}
	h_{i,a}(t) = {\left(\mathbf{w}^r_i(t)\right)} ^*\mathbf{H}_{i,a}(t)\mathbf{w}^t_i(t).
\end{equation}

Meanwhile, consider another vehicle $k$ that communicates with BS $\hat a$ over the uplink. The interfered channel gain caused by vehicle $k$ to BS $a$ at period $t$ is given by
\begin{equation}
	\label{h_ij}
	\widetilde{h}^{k,\hat a}_{i,a}(t) = {\left(\mathbf{w}^r_i(t)\right)} ^*\mathbf{H}_{k,a}(t)\mathbf{w}^t_{k}(t).
\end{equation}

Each vehicle is assumed to transmit with power $P_v$.  
The association relationship between vehicles and BSs is represented by the binary variable $\beta_{ia}(t)$, where $\beta_{ia}(t)=1$ if vehicle $i$ is associated with BS $a$ at time $t$, and $\beta_{ia}(t)=0$ otherwise.  
Given the bandwidth $W$ and the thermal noise power density $N_o$, the Signal-to-Interference-plus-Noise Ratio (SINR) at BS $a$ when serving vehicle $i$ is defined as 
\begin{equation}
	\label{I_ia}
	\text{SINR}_{i}^a(t) = \frac{P_v|h_{i,a}(t)|^2}{\Big|\sum_{k \in (\mathbb{U}(t) \backslash i)} \sum\limits_{\hat a \in \mathbb{B}}  \sqrt{P_v} \widetilde{h}^{k,\hat a}_{i,a}(t) \beta_{k\hat a}(t) \Big| ^ 2 + N_oW}.
\end{equation}

Based on (\ref{h_ij}) and (\ref{I_ia}), the total network transmission rate at period $t$ can then be expressed as
\begin{equation}
	\begin{aligned}
		r(t) = \sum_{i \in \mathbb{U}(t)}\sum_{a \in \mathbb{B}}  \beta_{ia}(t)W\log_2(1 + \text{SINR}_{i}^a(t)).
	\end{aligned}
\end{equation}

The objective of this study is to determine the optimal user association as well as the beamforming vectors and combining vectors that maximize the total instantaneous transmission rate of all vehicles.  
Let $\boldsymbol{\beta}(t)$, $\mathbf{W}^t(t)$, and $\mathbf{W}^r(t)$ denote the decision variables of the optimization problem.  
Specifically, $\boldsymbol{\beta}(t) = \{\beta_{ia}(t) \mid i \in \mathbb{U}(t), a \in \mathbb{B}\}$ represents the association vector between vehicles and BSs at time $t$.  
$\mathbf{W}^t(t) = \{\mathbf{w}^t_i(t) \mid i \in \mathbb{U}(t)\}$ and $\mathbf{W}^r(t) = \{\mathbf{w}^r_i(t) \mid i \in \mathbb{U}(t)\}$ denote the beamforming vectors and combining vectors, respectively.
\begin{equation}
	\label{formulate_semi_ACK_UCB}
	\begin{aligned}
		&\max_{\boldsymbol{\beta}(t), \mathbf{W}^t(t), \mathbf{W}^r(t)} \quad \sum_{i \in \mathbb{U}(t)}\sum_{a \in \mathbb{B}}  \beta_{ia}(t)W\log_2\left(1 + \text{SINR}_{i}^a(t)\right)\\
		&\begin{array}{r@{\quad}r@{}l@{\quad}l}
			s.t. &\sum\limits_{a \in \mathbb{B}}\beta_{ia}(t) = 1, \mathbf{w}^t_i(t) \in \mathcal{W}_T, \mathbf{w}^r_i(t) \in \mathcal{W}_R, \forall i \in \mathbb{U}(t).\\
		\end{array}
	\end{aligned}
\end{equation}
The first constraint ensures that each vehicle can connect to only one BS, while a BS can serve multiple vehicles.
In this work, we focus on the association $\boldsymbol{\beta}(t)$ and the beamforming vector $\mathbf{W}^t(t)$.
BSs are typically equipped with massive antenna arrays and powerful signal processing capabilities, which enable them to obtain near-optimal combining vectors.

\section{beam-aware kernelized upper confidence bound Algorithm}
\label{section BKC-UCB}

This section begins with an overview of the proposed BKC-UCB algorithm.
We then describe its three main steps in detail and derive an upper bound on the cumulative external regret.

\subsection{Framework of BKC-UCB Algorithm}
\begin{figure}[htpb]
	\centering\includegraphics[scale=0.39]{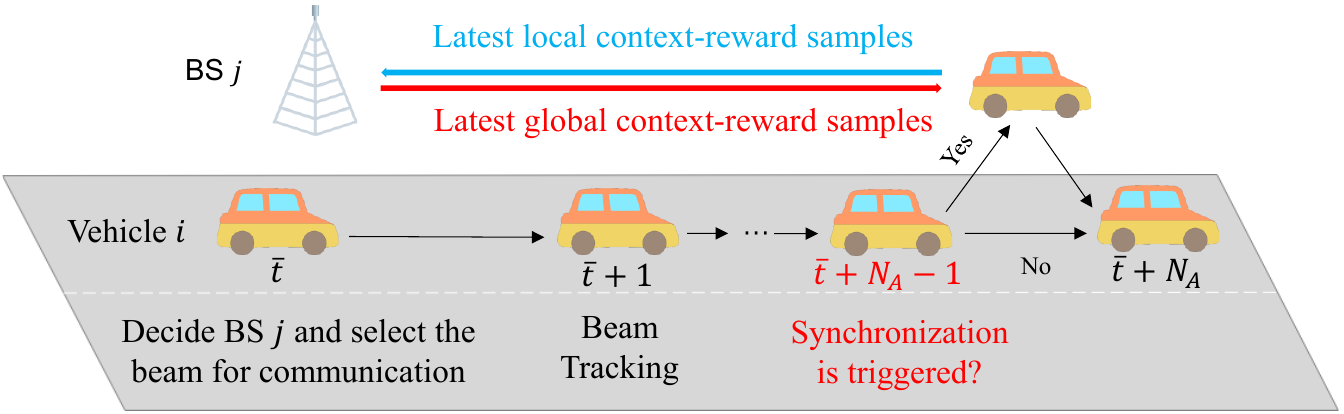}
	\caption{Architecture of the BKC-UCB algorithm during periods $[\bar t, \bar t + N_A - 1]$.}
	\label{algorithm_structure}
\end{figure}
Consider a vehicle $i$ at period $t \in [\bar t, \bar t + N_A - 1]$.
User association decisions are made at period $t=\bar t$ and updated every $N_A$ periods, since performing user association at every short-duration period would incur excessive handover overhead.  
Accordingly, the BS associated with vehicle $i$ is assumed to remain unchanged during periods $[\bar t+1,\, \bar t+N_A-1]$.
Beam tracking is executed at each period to adapt to fast-varying channel conditions.
Let $t_s = \bar t + N_A - 1$.
For periods $t \in [\bar t, t_s]$, the proposed algorithm operates as illustrated in Fig.~\ref{algorithm_structure}, which consists of the following three main steps.

\begin{itemize}
	\item \textbf{User Association Decision ($t = \bar t$)}:
	Using kernel methods, vehicle $i$ estimates the expected rewards for all BSs based on the latest global context-reward samples.
The BS with the highest estimated transmission rate is then selected for communication during the periods $[\bar t, t_s]$.

	\item \textbf{Beam Tracking ($t \in [\bar t, t_s]$)}:
	At each period, vehicle~$i$ selects the transmit beam by jointly determining the steering angle $\psi$ and the corresponding layer index $l$.  
	At period $t=\bar t$, instead of treating each beam independently, vehicle~$i$ selects the steering angle with the highest estimated transmission rate using the CMAB framework with kernel methods.  
	The layer index $l$ is determined according to the corresponding standard deviation of the estimated reward, where larger uncertainty leads to the selection of a wider beam for robust exploration.  
	Since the associated BS remains fixed within the short interval $[\bar t+1, t_s]$, the channel can be treated as approximately stationary, allowing a hierarchical beam tracking procedure to refine the steering direction and layer index over the binary-tree-structured codebook for $t \in [\bar t+1, t_s]$.

	\item \textbf{Synchronization ($t = t_s$)}:
	Synchronization between vehicle $i$ and the BS is triggered only after adequate exploration has been conducted since the last synchronization, reducing communication overhead while maintaining high estimation accuracy.

\end{itemize}

\subsection{User Association Decision}

At period $t=\bar t$, vehicle $i$ selects a BS $a_i(t)$ from the candidate set $\mathcal{A}_i(t)\subseteq \mathbb{B}$, which consists of BSs within a distance $R_{\max}$, as BSs beyond this range experience severe path loss and poor channel quality.  
BKC-UCB adopts the upper confidence bound (UCB) index \cite{lai1985asymptotically} for action selection to balance exploration and exploitation.  
For each candidate BS $a\in\mathcal{A}_i(t)$ with context $\mathbf{x}_i^a(t)$, let $\hat{\mu}_i(\mathbf{x}_i^a(t))$ and $\hat{\sigma}_i(\mathbf{x}_i^a(t))$ denote the expectation and standard deviation of the estimated reward, respectively.  
The BS with the highest UCB index is selected for communication, where coefficient $\alpha$ controls the balance between exploration and exploitation.
\begin{equation}
	\label{arm_selection}
	\begin{aligned}
		a_i(t) = {\arg\max}_{a \in \mathcal{A}_i(t)} \hat{\mu}_i(\mathbf x_i^a(t)) + \alpha\hat{\sigma}_i(\mathbf x_i^a(t)).
	\end{aligned}
\end{equation}

Let $d$ indicate the dimension of the context.
The context vector for vehicle $i$ when selecting BS $a = a_i(t)$ is defined as 
$\mathbf{x}_i(a_i(t)) = (a,\theta_i^a(t), L_i^a(t), f_i^a(t), N_i^a(t), \Delta\psi_i(t)) \in \mathbb{R}^{d\times1}$, 
which captures the key factors affecting the transmission performance, detailed as follows.
\begin{itemize}
	\item $a=a_i(t)$: Index of the selected BS, i.e., an arm.
	\item $\theta_i^a(t)$: Angle between the positive x-axis and the vector from arm $a$ to vehicle $i$.
	\item $L_i^a(t)$: Euclidean distance from vehicle $i$ to BS $a$.
	\item $f_i^a(t)$: Maximum Doppler spread, given by $f_i^a(t)=\frac{v_i^a(t)}{\lambda_f}$, where $v_i^a(t)$ denotes the tangential velocity component of the vehicle relative to BS $a$, and $\lambda_f$ is the signal wavelength.
	\item $N_i^a(t)$: Number of concurrent transmissions of BS $a$ for the last period of vehicle $i$ communicating with BS $a$.
	\item $\Delta\psi_i(t)$: Bias between the selected transmit beam steering angle and the geometric angle of the dominant LOS path between vehicle $i$ and BS $a$.
\end{itemize}

Let $\mathbf S^a_i(t) \in \mathbb{R}^{|\mathbf S_i(t)| \times d}$ and $\mathbf R^a_i(t) \in \mathbb{R}^{|\mathbf S_i(t)| \times 1}$ denote the sets of available contexts and the corresponding rewards of vehicle $i$ associated with BS $a$ up to period $t$, respectively.  
Since the reward function is nonlinear with respect to the context, each context is implicitly mapped by a feature map $\phi$ into the RKHS, where a linear relationship between the mapped context and the reward can be established.  
The expectation and standard deviation of the estimated reward for a given context can be obtained using a specific kernel function $\kappa$ without knowledge of the feature mapping $\phi$ \cite{valko2013finite}.
\begin{equation}
	\label{estimated_reward_confidence}
	\begin{aligned}
		\hat\mu_i(\mathbf x_i^a(t),\kappa) &= \mathbf{K}_{\mathbf x_i^a(t), \mathbf S^a_i(t)}^\top(\mathbf{K}_{\mathbf S^a_i(t),\mathbf S^a_i(t)} + \lambda_k\mathbf I)^{-1}\mathbf R^a_i(t),\\
		\hat\sigma_i(\mathbf x_i^a(t),\kappa) &= \lambda_k^{-1/2} \Big(\kappa(\mathbf x_i^a(t),\mathbf x_i^a(t)) - \mathbf{K}_{\mathbf x_i^a(t), \mathbf S^a_i(t)}^\top\times\\
		&(\mathbf{K}_{\mathbf S^a_i(t),\mathbf S^a_i(t)} + \lambda_k\mathbf I)^{-1}\mathbf{K}_{\mathbf x_i^a(t), \mathbf S^a_i(t)}\Big)^{\frac 12},
	\end{aligned}
\end{equation}
where,
\begin{equation}
	\label{kernel_phi}
	\nonumber
	\begin{split}
		\mathbf{K}_{\mathbf x_i^a(t),\mathbf S^a_i(t)}
		&= \big[\,\kappa\big(\mathbf x_i^a(t),\hat{\mathbf x}_{\hat t,i}^a\big)\,\big]_{\substack{\hat{\mathbf x}_{\hat t,i}^a\in\mathbf S^a_i(t)\\ \mathbf x_i^a(t)\notin\mathbf S^a_i(t)}}\in \mathbb{R}^{|\mathbf S^a_i(t)|\times 1},\\
		\mathbf{K}_{\mathbf S^a_i(t),\mathbf S^a_i(t)}
		&= \big[\,\kappa\big(\bar{\mathbf x}_{t,i}^a,\hat{\mathbf x}_{\hat t,i}^a\big)\,\big]_{\substack{\bar{\mathbf x}_{t,i}^a\in\mathbf S^a_i(t)\\ \hat{\mathbf x}_{\hat t,i}^a\in\mathbf S^a_i(t)}}\in \mathbb{R}^{|\mathbf S^a_i(t)|\times|\mathbf S^a_i(t)|},\\
		\kappa\big(\mathbf x_i^a(t), \hat{\mathbf{x}}_{\hat i}^a(\hat t)\big) &:= \phi\big(\mathbf x_i^a(t)\big)^\top \phi\big(\hat{\mathbf{x}}_{\hat i}^a(\hat t)\big)_{\substack{\mathbf x_i^a(t), \hat{\mathbf{x}}_{\hat i}^a(\hat t) \in \mathbb{R}^d}}.
	\end{split} 
\end{equation}
Here, $\lambda_k$ denotes the regularization parameter.
The kernel function $\kappa\big(\mathbf x_i^a(t), \hat{\mathbf{x}}_{\hat i}^a(\hat t)\big)$ evaluates the similarity between two contexts $\mathbf x_i^a(t)$ and $\hat{\mathbf{x}}_{\hat i}^a(\hat t)$.
The transmission rate is determined by blockage, path loss, Doppler spread, interference, and beamforming gain.
Therefore, evaluating the similarity of each element individually is more accurate than relying on a single kernel function. 
Accordingly, we design the kernel functions by explicitly accounting for the characteristics of mmWave vehicular channels.

The probability of the signal being blocked by buildings is proportional to $\cos\Delta \theta$ given a small $\Delta \theta = | \theta_i^a(t)-\hat{\theta}_{\hat i}^a(\hat t)|$ \cite{8114332}, so the similarity regarding the angle is modeled as
\begin{equation}
	\label{k_theta}
	k_{\theta}(\theta_i^a(t),\hat{\theta}_{\hat i}^a(\hat t)) =
	\begin{cases}
		\cos\Delta \theta, & \Delta\theta < \frac{\pi}{2},\\
		0, & \text{otherwise}.
	\end{cases}
\end{equation}

The distance to the BS also affects blockage probability and path loss.  
Let $\Delta L = |L_i^a(t)-\hat{L}_{\hat i}^a(\hat t)|$.  
A large $\Delta L$ implies that two contexts may experience distinct blockage conditions.  
To capture the similarity in blockage and path loss with respect to distance, the Gaussian kernel is used to model the nonlinear characteristics \cite{shawe2004kernel}.
\begin{equation}
	\label{k_L}
	k_L(L_i^a(t),\hat{L}_{\hat i}^a(\hat t)) = 
	\exp\!\left(-\Delta L^2/(2\sigma_L^2)\right).
\end{equation}
The exponential kernel, which is less sensitive to variations, is used to capture the similarity in Doppler spread \cite{paclik2000road}.
\begin{equation*}
	k_f(f_i^a(t),\hat{f}_{\hat i}^a(\hat t)) =
	\exp\!\left(-|f_i^a(t)-\hat{f}_{\hat i}^a(\hat t)|/\sigma_f\right).
\end{equation*}
The interference is dominated by the number of concurrent transmissions, and the similarity of two contexts in interference by deploying the triangular kernel \cite{genton2001classes}
\begin{equation*}
	k_N(N_i^a(t),\hat{N}_{\hat i}^{\hat a}(t)) = (1-|N_i^a(t)-\hat{N}_{\hat i}^{\hat a}(t)|/\sigma_N)^+.
\end{equation*}
Then, the kernel function for user association is defined as
\begin{equation}
	\label{kernel_proposed}
	\begin{aligned}
		\kappa = k_{\theta}\, k_L\, k_f\, k_N.
	\end{aligned}
\end{equation}
The kernel function between contexts associated with different BSs is set to zero, as their reward distributions for different BSs are considered independent.

For beam tracking, the bias $\Delta\psi_i(t)$ is additionally incorporated into the context to capture correlations among different beams.  
The similarity between steering angles is modeled by the Gaussian kernel
\begin{equation*}
	k_\psi(\psi_i^a(t),\hat{\psi}_{\hat i}^a(\hat t)) =
	\exp\!\left(-|\psi_i^a(t)-\hat{\psi}_{\hat i}^a(\hat t)|^2/(2\sigma_\psi^2)\right).
\end{equation*}
Extending (\ref{kernel_proposed}) by explicitly accounting for beam similarity, the kernel function for beam tracking is defined as
\begin{equation}
	\label{kernel_BT}
	\kappa^{BT} = k_{\theta}\, k_L\, k_f\, k_N\, k_\psi.
\end{equation}

The product of valid kernel functions is also a valid kernel function \cite{DK-UCB}, so both (\ref{kernel_proposed}) and (\ref{kernel_BT}) are valid kernels.

\subsection{Beam Tracking}
In a hierarchical binary-tree structured codebook, beam selection is typically performed via hierarchical search.  
For a beam represented by a parent node with steering angle $\bar{\psi}^a_i(t)$ and layer index $\bar l^a_i(t)$, the transceiver evaluates the channel gains of its two child nodes, and the one ${\psi}^a_i(t)$ with the higher estimated channel gain with layer $l^a_i(t)$ is selected for communication, i.e.,
\begin{equation}
	\label{determine_beam}
	\begin{aligned}
		{\psi}^a_i(t) &= \arg\max_{\psi' \in \{\bar{\psi}^a_i(t) \pm w(\bar{\psi}^a_i(t),l^a_i(t))/2\}} \hat r(\psi', l^a_i(t)),\\
		l^a_i(t) &= \min(\bar l^a_i(t) + 1,L_m).
	\end{aligned}
\end{equation}
Here $w(\psi,l)$ and $\hat r(\psi,l)$ denote the beamwidth and the transmission rate of the beam with layer $l$ and steering angle $\psi$.

At period \( t=\bar t \), user association is performed, and the associated BS may change.
Therefore, the beam search needs to restart from a new parent node with steering angle $\bar\psi_i^a(t)$ and layer $\bar l_i^a(t)$ for the new associated BS.
Let \( \{ \Delta\psi_n \}_{n=1}^{N_{\Delta_\psi}} \) be the set of biases between the steering angles of all available beams and the geometric angle of the dominant LOS path.
The steering angle \( \bar\psi_i^a(t) \) with the highest estimated reward is selected for transmission.
The estimation is following (\ref{estimated_reward_confidence}), with the kernel function $\kappa$ specified as $\kappa^{BT}$ in (\ref{kernel_BT}).
\begin{equation}
	\label{parent_kernel_psi}
	\begin{aligned}
		\bar\psi^a_i(t) &= \Delta\psi_{n^*} + \rho_i^a(t), n^* = \arg\max_{n\in N_{\Delta_\psi}} \hat\mu_i(\mathbf x_{i,n}^a(t),\kappa^{BT}), \\
		\mathbf x_{i,n}^a(t) &= (a,\theta_i^a(t), L_i^a(t), f_i^a(t), N_i^a(t), \Delta\psi_n).
	\end{aligned}
\end{equation}
Where $\rho_i^a(t)$ denotes the geometric steering angle of the LOS path to BS $a$.
The layer index \( \bar l^a_i(t) \) is determined based on the standard deviation of the estimated reward $\hat\sigma_i(\mathbf x_{i,n}^a(t),\kappa^{BT})$, and a large $\hat\sigma_i(\mathbf x_{i,n}^a(t),\kappa^{BT})$ implies low estimation confidence and necessitates a low layer and a wide beam.
\begin{equation}
	\label{parent_kernel_l}
	\bar l^a_i(t)
	= \left\lceil L_m\big(1 - \lambda_k\left(\hat\sigma_i(\mathbf x_{i,n^*}^a(t),\kappa^{BT})\right)^2 \big) \right\rceil - 1.
\end{equation}

At period \( t \in [\bar t + 1, \bar t + N_A - 1] \), since the associated BS remains unchanged and the duration of a period is short, the hierarchical search continues from the parent node selected in the previous period.

\subsection{Synchronization}
To balance communication cost and estimation accuracy, a vehicle triggers synchronization only when sufficient exploration has been conducted.
Let $t^{i}_{syn}$ denote the period of the last synchronization for vehicle $i$.
The trigger-event for vehicle $i$ and BS $a$ at period $t = t_s$ is defined as,
\begin{equation}
	\label{synchronization_event}
	\begin{aligned}
		&\mathcal{U}_{t,i}(L) = \big\{(t - t^{i}_{syn})\times\\&\log\Big(\frac{\det(\mathbf I + \lambda_k^{-1}\mathbf K_{\mathbf S_i^a(t),\mathbf S_i^a(t)})}{\det(\mathbf I + \lambda_k^{-1}\mathbf K_{\mathbf S_i^a(t)\backslash \mathbf S_i^a(t^{i}_{syn}),\mathbf S_i^a(t)\backslash \mathbf S_i^a(t^{i}_{syn})})}\Big) > L \big\}.
	\end{aligned}
\end{equation}
Here, $L$ is a coefficient to achieve a trade-off between communication cost and estimation accuracy.
$\mathcal{U}_{t,i}(L)$ is true if vehicle $i$ has accumulated plenty of new context-reward samples with low similarity to the past contexts for BS $a$ since $t^{i}_{syn}$.
Let $\mathbf S_i(t) \in \mathbb{R}^{|\mathbf S_i(t)| \times d}$ and $\mathbf R_i(t) \in \mathbb{R}^{|\mathbf S_i(t)| \times 1}$ denote the sets of available contexts and the corresponding rewards of vehicle $i$ up to period $t$, respectively.  
Then, the proposed BKC-UCB algorithm is presented in Algorithm \ref{BKC-UCB_alg}.

\begin{algorithm}
	\caption{BKC-UCB at period $t \in [\bar t,\, t_s]$}
	\label{BKC-UCB_alg}
	
	\eIf{$t = \bar t$}{
		Determine the associated BS $a = a_i(t)$ by (\ref{estimated_reward_confidence})
		
		Determine $(\bar\psi_i^a(t), \bar l_i^a(t))$ using (\ref{parent_kernel_psi}) and (\ref{parent_kernel_l})
	}{
		Set $\bar\psi_i^a(t) = \psi_i^a(t-1)$ and $\bar l_i^a(t) = l_i^a(t-1)$
	}
	
	Determine the beam $\mathbf w(\psi_i^a(t), l_i^a(t))$ by (\ref{determine_beam})
	
	\If{$t = t_s$ and $\mathcal{U}_{t,i}(L)$ is TRUE}{
		Send $\mathbf S_i(t)\setminus \mathbf S_i(t^i_{{syn}})$ and $\mathbf R_i(t)\setminus \mathbf R_i(t^i_{{syn}})$ to BS $a$
		
		Receive $\bigcup_{k\neq i}^{\mathbb{U}(t)} \mathbf S_k(t^k_{{syn}})\setminus \mathbf S_i(t^i_{{syn}})$ and rewards
		
		$t^i_{{syn}} = t$
	}
\end{algorithm}


\subsection{Regret Analysis}

When the actions of all other agents are fixed, vehicle~$i$ aims to minimize its external regret, which quantifies the gap between the expected reward of its current strategy and that of the best action at each period \cite{6953296}.
Let $\mathbf{z}^-_{i,t}$ denote the associations and beamforming vectors of all vehicles except vehicle $i$ at period $t$. Denote $\beta_i^*(t)$ and $\mathbf{w}_i^*(t)$ as the optimal association and beamforming vector for vehicle $i$ at period $t$, respectively.
The cumulative external regret is
\begin{equation}
	\begin{aligned}
		\mathcal{R}_{i}(T)
		=
		\sum_{t=1}^{T}
		\Big[
		R\big(\beta_{i}^*(t), \mathbf{w}^*_i(t), \mathbf{z}^-_{i,t}\big)
		-
		R\big(\beta_{i}(t), \mathbf{w}_i(t), \mathbf{z}^-_{i,t}\big)
		\Big],
	\end{aligned}
\end{equation}
where $R\big(\beta_{i}(t), \mathbf{w}_i(t), \mathbf{z}^-_{i,t}\big)$ denote the reward for vehicle $i$ with the associated BS $\beta_{i}(t)$ and the beamforming vector $\mathbf{w}_i(t)$ selected by BKC-UCB under $\mathbf{z}^-_{i,t}$.
Then, we have
\begin{equation}
	\begin{aligned}
		\mathcal{R}_{i}(T)
		&=
		\mathcal{R}_{i}^{\mathrm{1}}(T)
		+
		\mathcal{R}_{i}^{\mathrm{2}}(T),\\
		\mathcal{R}_{i}^{\mathrm{1}}(T)
		&\triangleq
		\sum_{t=1}^{T}
		\Big[
		R\big(\beta_{i}^*(t), \mathbf{w}_i(t), \mathbf{z}^-_{i,t}\big)
		-
		R\big(\beta_{i}(t), \mathbf{w}_i(t), \mathbf{z}^-_{i,t}\big)
		\Big],\\
		\mathcal{R}_{i}^{\mathrm{2}}(T)
		&\triangleq
		\sum_{t=1}^{T}
		\Big[
		R\big(\beta_{i}^*(t), \mathbf{w}^*_i(t), \mathbf{z}^-_{i,t}\big)
		-
		R\big(\beta_{i}^*(t), \mathbf{w}_i(t), \mathbf{z}^-_{i,t}\big)
		\Big].
	\end{aligned}
\end{equation}
$\mathcal{R}_{i}^{\mathrm{1}}(T)$ and $\mathcal{R}_{i}^{\mathrm{2}}(T)$ denote the external regrets associated with user association under the beamforming vectors selected by BKC-UCB, and with beamforming under the optimal user association, respectively.

Since the kernel functions defined in~(\ref{kernel_proposed}) and~(\ref{kernel_BT}) deploy the Gaussian kernel function, their maximum information gains can be both bounded by $ \gamma_T = \mathcal{O}(\log^2 T)$ \cite{seeger2008information,ACK-UCB}.
Then, $\mathcal{R}_{i}^{\mathrm{1}}(T)$ can be bounded by the following lemma.

\begin{lemma}[\cite{wang2019distributed,li2022communication} ]
	\label{lemma_BKC-UCB}
	With threshold $D = T/\gamma_{T}$ and the norm of the optimal reward parameter vector $\|\theta_*\|$ is bounded in the RKHS, $\alpha = \sqrt{\lambda_k}\|\theta_*\| + R\sqrt{4\log T/\delta + 2\log\det(\mathbf I + \lambda_k^{-1}\mathbf K_{\mathbf S_i(t),\mathbf S_i(t)})}$, $\mathcal{R}^1_{i}(T)$ is upper bounded by $\mathcal{O}(\sqrt{T}(\|\theta_*\|\sqrt{\gamma_{T}}+ \gamma_{T}))$ with probability at least $1 - \delta$.
\end{lemma}

Regarding $\mathcal{R}_{i}^{\mathrm{2}}(T)$, the worst case occurs when the vehicle lacks any prior information about beamforming after each association decision and restarts the hierarchical beam search from the first layer.
In such cases, the external regret $\mathcal{R}_{i}^{\mathrm{2}}(T)$ can be bounded by the cost of performing a full hierarchical beam search after each association update.

\section{Numerical Results}

The evaluation considers a dense urban region in Tokyo, Japan, with an area of approximately $1268\text{m} \times 1206\text{m}$, delimited by $(35.709241^\circ\text{N}, 139.751112^\circ\text{E})$ and $(35.719924^\circ\text{N}, 139.764963^\circ\text{E})$.
Using the urban topology data (roads, traffic signals, and buildings) extracted from OpenStreetMap~\cite{OpenStreetMap}, vehicle trajectories are generated by the SUMO simulator~\cite{SUMO2018} under realistic traffic conditions.
The ray-tracing-based clustered delay line (CDL) channel model~\cite{3gpp2018study} is used to model the mmWave vehicular channel, considering the static blockages caused by the buildings and the Doppler spread.
The deployment of BSs follows real-world measurements from the OpenCellID database~\cite{opencellid}, and 40 BS locations within the considered area are randomly selected for the simulation.
The worst case is assumed that all vehicles share the same bandwidth, leading to both intra-cell and inter-cell interference.
The arrival rate is 0.4$~\mathrm{s}^{-1}$, and the corresponding expected vehicle density is 24.07 (1/km$^2$).

The offline centralized Worst Connection Swapping (WCS) algorithm~\cite{8677293} and online distributed DK-UCB algorithm~\cite{DK-UCB} are considered as benchmarks.
WCS requires full CSI between all vehicles and BSs to estimate the transmission rates to determine the association and derive beamforming and combining vectors through SVD.
On the other hand, DK-UCB operates under a kernelized CMAB framework to estimate the transmission rate between vehicles and BSs, but beamforming is carried out under the assumption that CSI between each vehicle and its associated BS is available.
\begin{figure}[htpb]
	\centering\includegraphics[scale=0.55]{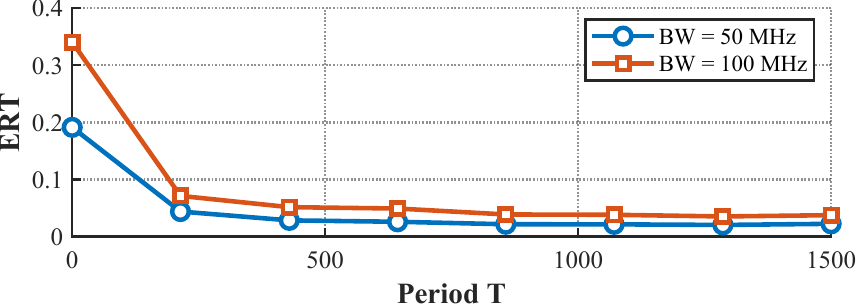}
	\caption{Cumulative external regret to time horizon $\mathbb{E}[\mathcal{R}_i(T)] / T$ (ERT).}
	\label{Fig_1}
\end{figure}

Given a transmit power of 30dBm, Fig.~\ref{Fig_1} shows the cumulative external regret to the time horizon $\mathbb{E}[\mathcal{R}_i(T)] / T$ of the proposed BKC-UCB algorithm with bandwidths of 50MHz and 100MHz.
The ERT starts at 0.34 and gradually decreases to 0.037 after the time period reaches 1500 under a bandwidth of 100 MHz.
In the initial stage, BKC-UCB has no prior knowledge of the network and therefore performs more exploration, which may result in pulling suboptimal arms.
As the learning process converges, BKC-UCB gradually acquires accurate network knowledge and shifts toward exploitation.
\begin{figure}[h]
	\centering\includegraphics[scale=0.5]{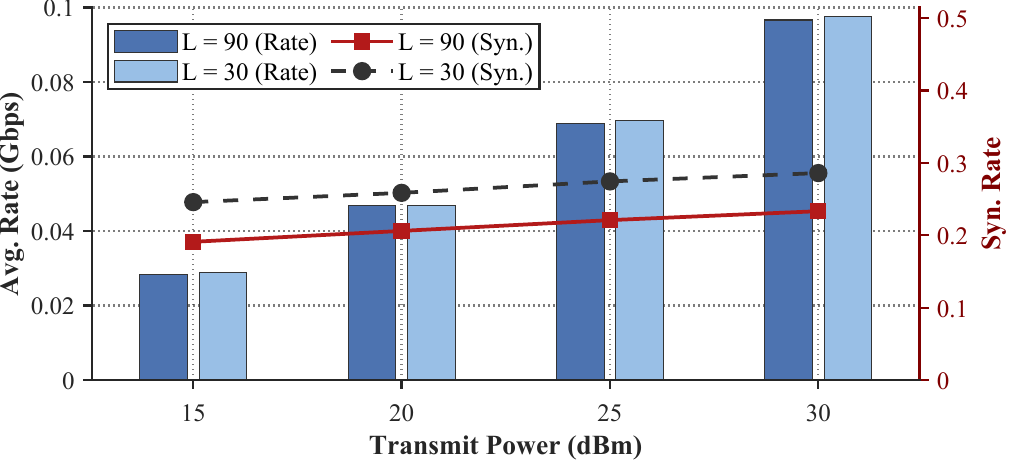}
	\caption{Average transmission rate (Avg. Rate) per vehicle and synchronization rate (Syn. Rate) against the transmit power.}
	\label{Fig_2}
\end{figure}

Given the bandwidth of 100MHz, Fig.~\ref{Fig_2} presents the average transmission rate over the periods $[1900,2900]$ against the transmit power and coefficient $L$.
The synchronization rate is defined as the ratio of synchronization periods to the total number of periods.  
With a transmit power of 25 dBm, BKC-UCB with $L=30$ achieves an average data rate that is 1.1\% higher than that of BKC-UCB with $L=90$, while incurring a 5.4\% increase in the synchronization rate.  
It demonstrates that the parameter $L$ governs a trade-off between communication overhead and reward estimation accuracy.
\begin{figure}[htpb]
	\centering\includegraphics[scale=0.55]{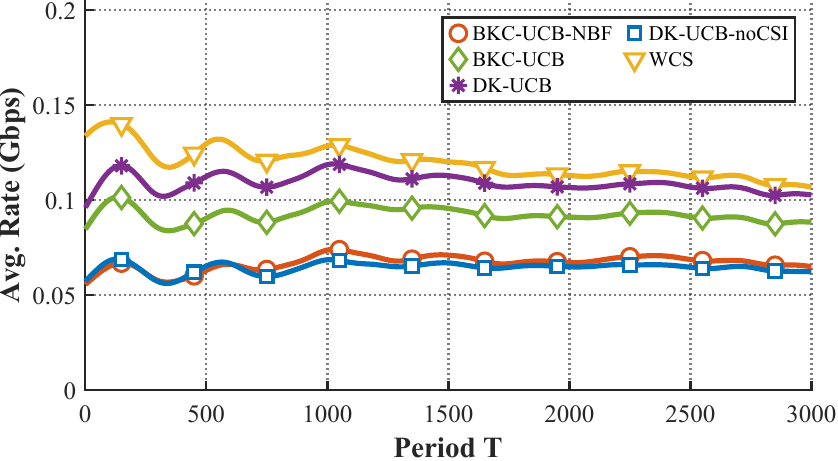}
	\caption{Average transmission rate per vehicle (Avg. Rate) against period $T$.}
	\label{Fig_3}
\end{figure}

Fig.~\ref{Fig_3} presents the expected transmission rate per vehicle given the bandwidth of 100MHz.
The expected transmission rate of BKC-UCB is 0.0914 Gbps over the first 500 periods, while it drops to 0.0895 Gbps during periods 2500-3000.
This trend can be attributed to the growth in vehicle density as time progresses, which results in more severe interference and consequently reduces the achievable transmission rate.
WCS outperforms BKC-UCB by 22.8\%, but WCS requires instantaneous full CSI in an offline manner, and BKC-UCB operates without any CSI.
Although DK-UCB outperforms BKC-UCB by 0.0169~Gbps, BKC-UCB achieves a 0.0273~Gbps higher expected transmission rate than DK-UCB without CSI.
This performance gap arises because DK-UCB requires the instantaneous CSI between each vehicle and its associated BS for beamforming, whereas BKC-UCB achieves competitive performance without relying on any CSI.
BKC-UCB achieves a 37.75\% higher expected transmission rate than the variant that restarts beam search from the first codebook layer, demonstrating its beamforming effectiveness.


%

\section{Conclusion}
This paper proposes the BKC-UCB algorithm for joint user association and beamforming in mmWave vehicular networks.
Leveraging kernel methods, the algorithm captures the nonlinear relationship between contexts and rewards, enabling transmission rate estimation and beam selection without requiring any CSI.
Also, BKC-UCB triggers information sharing only when significant exploration are conducted, accelerating convergence while keeping communication overhead manageable.

\bibliographystyle{IEEEtran}
\bibliography{reference}

@inproceedings{valko2013finite,
	title={Finite-time Analysis of Kernelised Contextual Bandits},
	author={Valko, Michal and Korda, Nathan and Munos, R{\'e}mi and Flaounas, Ilias and Cristianini, Nello},
	booktitle={Proceedings of the Twenty-Ninth Conference on Uncertainty in Artificial Intelligence},
	pages={654--663},
	year={2013}
}

@ARTICLE{8114332,
	author={Gupta, Abhishek K. and Andrews, Jeffrey G. and Heath, Robert W.},
	journal={IEEE Transactions on Wireless Communications}, 
	title={Macrodiversity in Cellular Networks With Random Blockages}, 
	year={2018},
	volume={17},
	number={2},
	pages={996-1010},
	doi={10.1109/TWC.2017.2773058}
	}

@article{shawe2004kernel,
	title={Kernel Methods for Pattern Analysis},
	author={Shawe-Taylor, J},
	journal={Cambridge University Press google schola},
	volume={2},
	pages={181--201},
	year={2004}
}

@article{paclik2000road,
	title={Road Sign Classification Using Laplace Kernel Classifier},
	author={Pacl{\i}k, Pavel and Novovi{\v{c}}ov{\'a}, J and Pudil, Pavel and Somol, Petr},
	journal={Pattern Recognition Letters},
	volume={21},
	number={13-14},
	pages={1165--1173},
	year={2000},
	publisher={Elsevier}
}

@article{genton2001classes,
	title={Classes of Kernels for Machine Learning: A Statistics Perspective},
	author={Genton, Marc G},
	journal={Journal of Machine Learning Research},
	volume={2},
	number={Dec},
	pages={299--312},
	year={2001}
}

@article{seeger2008information,
	title={Information Consistency of Nonparametric {Gaussian} Process Methods},
	author={Seeger, Matthias W and Kakade, Sham M and Foster, Dean P},
	journal={IEEE Transactions on Information Theory},
	volume={54},
	number={5},
	pages={2376--2382},
	year={2008},
	publisher={IEEE}
}

@techreport{3gpp2018study,
	title={Study on Channel Model for Frequencies from 0.5 to 100 {GHz} (Release 15)},
	author={3GPP Radio Access Network Working Group and others},
	year={2018},
	institution={3GPP TR 38.901}
}

@article{lai1985asymptotically,
	title={Asymptotically Efficient Adaptive Allocation Rules},
	author={Lai, Tze Leung and Robbins, Herbert and others},
	journal={Advances in Applied Mathematics},
	volume={6},
	number={1},
	pages={4--22},
	year={1985}
}

@article{li2020smart,
	title={Smart Vehicular Communication via {5G} M{mWaves}},
	author={Li, Xiaotong and Zhou, Ruiting and Zhang, Ying-Jun Angela and Jiao, Lei and Li, Zongpeng},
	journal={Computer Networks},
	volume={172},
	pages={107173},
	year={2020},
	publisher={Elsevier}
}

@ARTICLE{8472783,
	author={Sim, Gek Hong and Klos, Sabrina and Asadi, Arash and Klein, Anja and Hollick, Matthias},
	journal={IEEE/ACM Transactions on Networking}, 
	title={An Online Context-Aware Machine Learning Algorithm for {5G} m{mWave} Vehicular Communications}, 
	year={2018},
	volume={26},
	number={6},
	pages={2487-2500},
	doi={10.1109/TNET.2018.2869244}
	}

@article{CC-UCB,
	title={Contextual Bandits with Non-Stationary Correlated Rewards for User Association in MmWave Vehicular Networks},
	author={He, Xiaoyang and Huang, Xiaoxia and Li, Lanhua},
	journal={IEEE Transactions on Mobile Computing},
	year={2025},
	publisher={IEEE}
}

@article{wang2019distributed,
	title={Distributed Bandit Learning: Near-optimal Regret with Efficient Communication},
	author={Wang, Yuanhao and Hu, Jiachen and Chen, Xiaoyu and Wang, Liwei},
	journal={arXiv preprint arXiv:1904.06309},
	year={2019}
}

@article{li2022communication,
	title={Communication Efficient Distributed Learning for Kernelized Contextual Bandits},
	author={Li, Chuanhao and Wang, Huazheng and Wang, Mengdi and Wang, Hongning},
	journal={Advances in Neural Information Processing Systems},
	volume={35},
	pages={19773--19785},
	year={2022}
}

@ARTICLE{8677293,
	author={Alizadeh, Alireza and Vu, Mai},
	journal={IEEE Transactions on Wireless Communications}, 
	title={Load Balancing User Association in Millimeter Wave {MIMO} Networks}, 
	year={2019},
	volume={18},
	number={6},
	pages={2932-2945},
	doi={10.1109/TWC.2019.2906196}
	}

@misc{OpenStreetMap,
	author = {{OpenStreetMap Contributors}},
	title = {Planet Dump Retrieved from https://planet.osm.org },
	howpublished = "\url{ https://www.openstreetmap.org }",
	year = {2017},
}

@misc{opencellid,
	title        = {OpenCellID: The World's Largest Open Database of Cell Towers},
	author       = {{OpenCellID Community}},
	year         = {2025},
	howpublished = {\url{https://www.opencellid.org}},
	note         = {Accessed: 2024}
}

@ARTICLE{9779354,
	author={Li, Jing and Niu, Yong and Wu, Hao and Ai, Bo and Chen, Sheng and Feng, Zhiyong and Zhong, Zhangdui and Wang, Ning},
	journal={IEEE Communications Surveys $\&$ Tutorials}, 
	title={Mobility Support for Millimeter Wave Communications: Opportunities and Challenges}, 
	year={2022},
	volume={24},
	number={3},
	pages={1816-1842},
	doi={10.1109/COMST.2022.3176802}
	}

@ARTICLE{6497017,
	author={Ye, Qiaoyang and Rong, Beiyu and Chen, Yudong and Al-Shalash, Mazin and Caramanis, Constantine and Andrews, Jeffrey G.},
	journal={IEEE Transactions on Wireless Communications}, 
	title={User Association for Load Balancing in Heterogeneous Cellular Networks}, 
	year={2013},
	volume={12},
	number={6},
	pages={2706-2716},
	doi={10.1109/TWC.2013.040413.120676}
	}

@ARTICLE{6953296,
	author={Maghsudi, Setareh and Stańczak, Sławomir},
	journal={IEEE Transactions on Vehicular Technology}, 
	title={Joint Channel Selection and Power Control in Infrastructureless Wireless Networks: A Multiplayer Multiarmed Bandit Framework}, 
	year={2015},
	volume={64},
	number={10},
	pages={4565-4578},
	doi={10.1109/TVT.2014.2369425}
	}

@inproceedings{SUMO2018,
	title = {Microscopic Traffic Simulation using SUMO},
	author = {Pablo Alvarez Lopez and Michael Behrisch and Laura Bieker-Walz and Jakob Erdmann and Yun-Pang Fl{\"o}tter{\"o}d and Robert Hilbrich and Leonhard L{\"u}cken and Johannes Rummel and Peter Wagner and Evamarie Wie{\ss}ner},
	publisher = {IEEE},
	booktitle = {The 21st IEEE International Conference on Intelligent Transportation Systems},
	year = {2018},
	journal = {{IEEE Intelligent Transportation Systems Conference (ITSC)}},
	url = {https://elib.dlr.de/124092/}
}

@inproceedings{DK-UCB,
	title={Learning-Based User Association for M{mWave} Vehicular Networks with Kernelized Contextual Bandits},
	author={He, Xiaoyang and Huang, Xiaoxia},
	booktitle={2025 IEEE Wireless Communications and Networking Conference (WCNC)},
	pages={1--6},
	year={2025},
	organization={IEEE}
}

@article{al2022review,
	title={A review of the state of the art and future challenges of deep learning-based beamforming},
	author={Al Kassir, Haya and Zaharis, Zaharias D and Lazaridis, Pavlos I and Kantartzis, Nikolaos V and Yioultsis, Traianos V and Xenos, Thomas D},
	journal={{IEEE Access}},
	volume={10},
	pages={80869--80882},
	year={2022},
	publisher={IEEE}
}

@ARTICLE{1143830,
	author={Schmidt, R.},
	journal={IEEE Transactions on Antennas and Propagation}, 
	title={Multiple emitter location and signal parameter estimation}, 
	year={1986},
	volume={34},
	number={3},
	pages={276-280},
	doi={10.1109/TAP.1986.1143830}}

@article{xiao2016hierarchical,
	title={Hierarchical codebook design for beamforming training in millimeter-wave communication},
	author={Xiao, Zhenyu and He, Tong and Xia, Pengfei and Xia, Xiang-Gen},
	journal={IEEE Transactions on Wireless Communications},
	volume={15},
	number={5},
	pages={3380--3392},
	year={2016},
	publisher={IEEE}
}

@ARTICLE{ACK-UCB,
	author={He, Xiaoyang and Huang, Xiaoxia and Tsukada, Manabu},
	journal={IEEE Transactions on Wireless Communications}, 
	title={{ACK-UCB}: An Asynchronous Contextual Kernel-based Bandit Approach for User Association in MmWave Vehicular Networks}, 
	year={2026},
	volume={},
	number={},
	pages={1-1},
	doi={10.1109/TWC.2026.3658630}}

\end{document}